         \def\version{June 15, 2000}

\documentclass[reqno,12pt]{amsart}
\usepackage{amsmath}
\usepackage{amssymb}
\usepackage{euscript}

\usepackage{times,amssymb}
\renewcommand{\slshape}{\itshape}
\renewcommand{\sffamily}{\rmfamily}

\newfont{\scfont}{cmcsc10 scaled 1200}

\newfont{\bfit}{cmbxti10 scaled 1200}
\newcommand{\eps}{\varepsilon}
\newcommand{\supp}{{\operatorname {supp}\,}}
\newcommand{\prob}{{\operatorname {Prob}}}

\newcommand{\esssup}{{\operatorname {esssup}\mkern1mu}}
\newcommand{\essinf}{{\operatorname {essinf}\mkern1mu}}

\newcommand{\sgn}{{\operatorname{sgn}}}

\newcommand{\R}{\mathbb{R}}
\newcommand{\N}{\mathbb{N}}

\renewcommand{\P}{\mathbb{P}}
\newcommand{\Z}{\mathbb{Z}}

\newcommand{\E}{\mathbb{E}}
\newcommand{\1}{{\sf 1}}

\newcommand{\F}{{\mathcal F}}

\newcommand{\skrih}{{\mathcal H}}

\renewcommand{\L}{{\mathcal L}}

\newcommand{\DeltaD}{\Delta^{\mkern-2.3mu\operatorname{d}}\mkern0.2mu}
\newcommand{\lambdaD}{\lambda^{\operatorname{d}}}
\newcommand{\ED}{{\mathbb E}^{\operatorname{d}}}

\def\comment#1{}
\newtheoremstyle{thm}{1.5ex}{1.5ex}{\itshape\rmfamily}{}
{\bfseries\rmfamily}{}{2ex}{}

\newtheoremstyle{rem}{1.3ex}{1.3ex}{\rmfamily}{}
{\slshape\sffamily}
{}
{1.5ex}{}

\newenvironment{proofsect}[1]
{\vskip0.1cm\noindent{\sffamily\slshape #1.}}{\vspace{0.15cm}}

\theoremstyle{thm}
\newtheorem{theorem}{Theorem}[section]
\newtheorem{lemma}[theorem]{Lemma}
\newtheorem{proposition}[theorem]{Proposition}

\newtheorem*{Main Theorem}{Main Theorem.}
\newtheorem{cor}[theorem]{Corollary}

\theoremstyle{rem}
\newtheorem{remark}{{\slshape\sffamily Remark}}[]

\numberwithin{equation}{section}

\renewcommand{\section}{\secdef\sct\sect}
\newcommand{\sct}[2][default]{\refstepcounter{section}
\vspace{0.7cm}
\centerline{ 
\scfont\arabic{section}.\ #1}
\nopagebreak
\vspace{0.4cm}}
\newcommand{\sect}[1]{
\vspace{0.4cm}
\centerline{\large
\scfont #1}
\vspace{0.4cm}}

\renewcommand{\subsection}{\secdef \subsct\sbsect}
\newcommand{\subsct}[2][default]{\refstepcounter{subsection}
\vspace{0.2cm}
{\flushleft\bf\arabic{section}.\arabic{subsection}~\bf #1 }
\vspace{3mm}\nopagebreak}
\newcommand{\sbsect}[1]{\vspace{0.1cm}\noindent
{\bf #1}\vspace{0.1cm}}

\renewcommand{\subsubsection}{%
\secdef \subsubsect\sbsbsect}
\newcommand{\subsubsect}[2][default]{%
\refstepcounter{subsubsection}
\nopagebreak
\vspace{0.1\baselineskip}
\nopagebreak
{\flushleft
\sffamily\slshape
\arabic{section}.\arabic{subsection}.\arabic{subsubsection}
\ %
\sffamily #1\/.}\ }
\newcommand{\sbsbsect}[1]{\vspace{0.1cm}\noindent
{\bf #1}\ }

\begin{document}


\title[Screening in 1D parabolic Anderson model]
{\large Screening effect due to heavy lower tails in one-dimensional
parabolic Anderson model}

\author[Marek Biskup and Wolfgang K\"onig]{}
\maketitle

\thispagestyle{empty}
\vspace{0.2cm}
\centerline{\sc Marek Biskup$^1$\/
and Wolfgang K\"onig\small$^2$}
\vspace{0.8cm}
\centerline
{\em $^1$Microsoft Research, One Microsoft Way,
Redmond WA 98052, U.S.A.}
\centerline{{\em email: }{\sf biskup@microsoft.com}}
\centerline{\small and}
\centerline{\em $^2$Fachbereich Mathematik~MA7-5,
Technische Universit\"at Berlin,}
\centerline{\em
Stra\ss e des 17.~Juni~136, 10623 Berlin, Germany}
\centerline{{\em email: }{\sf koenig@math.tu-berlin.de}}
\vspace{0.5cm}
\centerline{\small(\version)}
\vspace{0.5cm}

\begin{quote}
\begin{quote}
{\footnotesize {\bf Abstract:} } \footnotesize
We consider the large-time behavior of
the solution $u\colon [0,\infty)\times\Z\to[0,\infty)$
to the parabolic Anderson problem $\partial_t u=\kappa\Delta u+\xi u\/\/$
with initial data $u(0,\cdot)=1$ and non-positive  finite i.i.d.\
potentials $(\xi(z))_{z\in\Z}$. Unlike in dimensions $d\ge2$,
the almost-sure decay rate of $u(t,0)$ as $t\to\infty$ is not
determined solely by the upper tails of $\xi(0)$;
too heavy lower tails of $\xi(0)$
accelerate the decay. The
interpretation is that sites $x$ with large negative $\xi(x)$ hamper the
mass flow and hence screen
off the influence of more favorable regions of the potential. The
phenomenon is unique to $d=1$. The result answers an open question
from our previous study \cite{BK00} of this model in general
dimension.
\end{quote}
\end{quote}

\vfill
\begin{tabular}{lp{13cm}}
\multicolumn{2}{l}
\footnotesize eigenvalues, screening effect.
\end{tabular}

\eject

\section{Introduction}

\subsection{Model and main aim}
\label{Mass}

\noindent
In a recent paper \cite{BK00}, we have studied the asymptotic
behavior of the solution $u(t,z)$ to the so-called parabolic
Anderson model for non-positive i.i.d.\ potentials. Here we answer
an open question concerning the almost-sure asymptotics of
$u(t,0)$ as $t\to\infty$ in dimension one for potentials lacking
the first logarithmic moment. Interestingly, a new phenomenon
arises: too heavy tails of the potential at $-\infty$ hamper the
mass flow to remote areas, thus rendering the more favorable
regions inaccessible. This effect is unique to $d=1$, since only
in one-dimensional topology particles are not able to bypass deep
broad valleys in the potential landscape.

The general model is defined as follows. Let $u\colon[0,\infty)
\times\Z^d\to[0,\infty)$ be the solution to the parabolic problem
\begin{equation}
\label{Anderson}
\begin{array}{rcll}
\displaystyle
\partial_t \,u(t,z)\!\!\! &=&\!\!\!\kappa \DeltaD u(t,z)+\xi(z)
u(t,z),\qquad &(t,z)\in(0,
\infty)\times \Z^d,\\
u(0,z)\!\!\!&=&\!\!\!1,&z\in\Z^d,
\end{array}
\end{equation}
where $\partial_t$ is the time derivative, $\kappa>0$ is the
diffusion constant, $\DeltaD$ is the discrete Laplacian on~$\Z^d$,
$[\DeltaD f](z)=\sum_{x\sim z}[f(x)-f(z)]$, and
$\xi=(\xi(z))_{z\in\Z^d}$ is an i.i.d.\ field. We use
$\langle\,\cdot\,\rangle$ to denote the expectation with respect
to $\xi$ and $\prob(\cdot)$ to denote the underlying probability
measure. One interpretation of the quantity $u(t,z)$ is the total
expected mass accumulated at time $t$ by a particle starting at
$z$ at time $0$ and diffusing through a random field of sources
(sites $x$ with $\xi(x)>0$) and sinks (sites $x$ with $\xi(x)<0$).
The references \cite{GM90}, \cite{CM94} and \cite{K00} provide
more explanation and other interpretations.

Besides \cite{BK00}, the large-$t$ behavior of the solution to
\eqref{Anderson} has extensively been studied (in general
dimension) for various other classes of distributions: see
\cite{GM90,GM98,GH99} for $\xi$ having the so-called
double-exponential upper tail, and \cite{GK98,GKM99} for a
continuous variant of \eqref{Anderson} with $\xi$ either Gaussian
or (smeared) Poissonian field.
The techniques used in these studies go back to the
pioneering work of Donsker and Varadhan \cite{DV75,DV79}; however,
there is also an intimate relation to Sznitman's method of
enlargement of obstacles \cite{S98}. We refer to \cite{K00} for a
comprehensive discussion of these relations and a unified
presentation of the above results. Henceforth, we shall focus on
the almost-sure behavior of $u(t,0)$ in the non-positive case,
i.e., $\xi\in[-\infty,0]^{\Z^d}$.

In dimensions $d\ge2$, the analysis in \cite{BK00} produced a
fairly complete picture. Indeed, interesting behavior occurs only
when $p=\prob(\xi(0)>-\infty)>p_{\text{c}}(d)$, the threshold for
site percolation on $\Z^d$, and when conditioned on the event that
the origin lies in the infinite cluster of sites $x$ with
$\xi(x)>-\infty$. Below and, provided there is no critical
percolation (which is rigorously known for $d=2$ \cite{R78} and
$d\ge19$ \cite{HS90}), also at $p_{\text{c}}(d)$, and also when
the origin lies in a finite cluster for $p>p_{\text{c}}(d)$, the
quantity $u(t,0)$ decays exponentially in $t$ with a
$\xi$-dependent rate.

In dimension $d=1$, we have $p_{\text{c}}(d)=1$, which
necessitated setting $\prob(\xi(0)=-\infty)=0$ in \cite{BK00}.
However, the latter condition was not sufficient because the
existence of the first logarithmic moment, i.e., $\langle
\log(-\xi(0)\vee1)\rangle<\infty$, also had to be assumed in order
to establish an asymptotics analogous to the supercritical case in
$d\ge2$. In particular, two intriguing questions remained
unanswered:
\begin{enumerate}
\item[$\bullet$] Is $\langle \log(-\xi(0)\vee1)\rangle<\infty$
optimal in the sense that $\langle
\log(-\xi(0)\vee1)\rangle=\infty$ implies a strictly different
asymptotic behavior of $u(t,0)$?
\item[$\bullet$] What is the precise decay rate when the finiteness
of $\langle \log(-\xi(0)\vee1)\rangle$ is robustly violated
(keeping however the restriction to ``no atom at $-\infty$'')?
\end{enumerate}

In this paper we give answers to these questions under mild
regularity conditions on the lower tail of the distribution of
$\xi$. In particular, we show that $\langle
\log(-\xi(0)\vee1)\rangle<\infty$ is only marginally non-optimal
for the behavior described in \cite{BK00}, see Remark~\ref{rem2}
after Theorem~\ref{mainres}. As it turns out, the decay of
$u(t,0)$ is determined solely by upper {\it and\/} lower tails of
$\prob(\xi(0)\in\cdot)$. The reason why the intermediate part of
the distribution does not play any role is that these tails give
rise to two dominant and mutually competing mechanisms
(field-shape optimization in the upper tail {\sl versus} screening
effect in the lower tail) whose balancing determines the decay
rate. See Subsection~\ref{heuristics} for more precise heuristic
explanation.

\subsection{Our assumptions}
\label{ass}

\noindent
We proceed by stating precisely the needed assumptions, both on
upper and lower tails of $\xi(0)$.  First we restrict ourselves to
dimension $d=1$ for the sequel of this paper. In accord with
\cite{BK00}, we consider the distributions with the upper tail of
the form
\begin{equation}
\label{perfect}
\prob\bigl(\xi(0)\ge-x\bigr)= \exp\bigl\{-
x^{-\frac\gamma{1-\gamma}+o(1)}\bigr\},\qquad x\downarrow0,
\end{equation}
for some $\gamma\in[0,1)$. However, instead of the distribution
function, it is more convenient to work with the cumulant
generating function
\begin{equation}
\label{Hdef}
H(\ell)=\log\langle e^{\ell\xi(0)}\rangle,\qquad \ell\ge0.
\end{equation}
The regime in \eqref{perfect} corresponds to the behavior
$H(\ell)=-\ell^{\gamma+o(1)}$ as $\ell\to\infty$.

\vspace{0.1cm}
\noindent{\bf Assumption~(H).\/}
{\it Let\/ $\esssup \xi(0)=0$ and suppose there are constants\/
$A>0$ and\/ $\gamma\in[0,1)$, and a positive increasing function\/
$t\mapsto\alpha_t$ such that
\begin{equation}
\label{mainass}
\lim_{t\to\infty}\frac{\alpha_t^{3}}{t}H\Bigl(\frac
t{\alpha_t} y\Bigr)=-A y^\gamma,\qquad y>0.
\end{equation}
}
\vspace{0.1cm}

The limit in \eqref{mainass} is necessarily uniform on compact
sets in $(0,\infty)$, the pair $(A,\alpha_t)$ is unique up to a
scaling transformation. Moreover, $t\mapsto\alpha_t$ is regularly
varying and $\alpha_t=t^{\nu+o(1)}$ as $t\to\infty$ where
$\nu=(1-\gamma)/(3-\gamma)\in(0,1/3]$. In particular,
$t/\alpha_t\to\infty$, i.e., Assumption~(H) indeed controls the
upper tails of $\xi(0)$. We say that $H$ {\it is in the
$\gamma$-class\/} if \eqref{mainass} holds.

Next we formulate our assumption on the lower tails of $\xi(0)$ at
$\essinf\xi(0)=-\infty$. As the opposite case has already been
handled in \cite{BK00}, we shall focus on the case where
$\log(-\xi(0)\vee1)$ is not integrable. Central to our attention
are lower tails of the form
\begin{equation}
\label{lowertails}
\prob\bigl(\log(-\xi(0)\vee1)>x\bigr)=x^{-\zeta+o(1)},\qquad x\to\infty,
\end{equation}
with some $\zeta\in[0,1]$. In terms of the modified cumulant
generating function
\begin{equation}
\label{G}
G(\ell)=-\log\bigl\langle (-\xi(0)\vee1)^{-1/\ell}\bigr\rangle,
\qquad \ell> 0,
\end{equation}
the behavior \eqref{lowertails} roughly corresponds to
$G(\ell)=\ell^{-\zeta+o(1)}$ as $\ell\to \infty$. Note that $G$ is
positive and decreasing since $\essinf\xi(0)<-1$. The following is
a weak regularity condition for $G$ at infinity.

\vspace{0.1cm}
\noindent{\bf Assumption~(G).\/}
{\it Let\/ $\langle\log(-\xi(0)\vee1)\rangle=\infty$ but\/
$\prob(\xi(0)=-\infty)=0$. Suppose that for each\/
$\eta\in(0,1)$\/ there is a function\/ $\widetilde G_\eta
\colon(0,\infty)\to(0,\infty)$ with the following properties:
\begin{enumerate}
\item[(i)] $\widetilde G_\eta(\ell)\le G(\ell)^{\eta+o(1)}$
as $\ell\to\infty$.
\item[(ii)]
$\ell\mapsto1/\widetilde G_\eta(\ell)$ is increasing  and concave
for $\ell$ large enough.
\item[(iii)]
The random variable $1/\widetilde G_\eta(\log(-\xi(0)\vee1))$ has
the first moment.
\end{enumerate}
}
\vspace{0.1cm}

\begin{remark}
\label{rem1}
As it turns out, Assumption~(G) is needed only for the proof of
the lower bound in our main result (see Theorem~\ref{mainres}
below); the upper bound requires no assumptions at all. The role
of Assumption~(G) and particularly of its part (i) is the
following: Abbreviate $Y=\log(-\xi(0)\vee1)$ and note that, for
any $\delta\in(0,1]$, $G(\ell)\le\langle Y^\delta\rangle
\ell^{-\delta}$. Therefore, $G(\ell)\le\ell^{-\zeta_*+o(1)}$ where
$\zeta_*=\sup\{\delta\ge0\colon\langle Y^\delta\rangle<\infty\}$.
However, a lower bound of the same (even asymptotic) form requires
some regularity of $\ell\mapsto G(\ell)$ as $\ell\to\infty$, which
is the essence of (i--iii).
\end{remark}

\begin{remark}
\label{rem1'}
In the view of Remark~\ref{rem1}, it is immediate that
Assumption~(G) holds for $G(\ell)= \ell^{-\zeta+o(1)}$ with some
$\zeta\in(0,1]$. The reason why we prefer the above (little
cumbersome) setting as opposed to simple regularity of $G$ is that
many cases with $G(\ell)=\ell^{o(1)}$ are automatically  included.
Indeed, consider the following example: Let $\theta>0$ and
$\prob(\log(-\xi(0)\vee1)\in dx)\sim C/[x\log^{1+\theta}(x)]dx$ as
$x\to\infty$, where $C$ is the normalizing constant. Then
$G(\ell)\sim C'(\log\ell)^{-\theta}$ and Assumption~(G) holds with
$\widetilde G_\eta(\ell)=G(\ell)[\log\log(\ell\vee
e)]^{1+\theta'}$ for any $\eta<1$ and any $\theta'>0$.
\end{remark}

\subsection{Main result}
\label{result}

\noindent
We begin by defining the scale function of the almost-sure
asymptotics:
\begin{equation}
\label{b_t}
\frac{b_t}{\alpha_{b_t}^2}=-\log G(t),\qquad t>0.
\end{equation}
In other words, $t\mapsto b_t$ is the inverse of the function
$t\mapsto t\alpha_t^{-2}$ (which we may and shall assume to be strictly
increasing),
evaluated at $-\log G(t)$.
Note that, since $\lim_{\ell\to\infty}G(\ell)=0$, we have $b_t\to\infty$ as
$t\to\infty$. If $G(\ell)=\ell^{-\zeta+o(1)}$ as $\ell\to\infty$ for some
$\zeta\in(0,1]$, then $\alpha_{b_t}^2= \zeta^{\beta}(\log t)^{\beta+o(1)}$,
where $\beta=2\nu/(1-2\nu)=2(1-\gamma)/(1-3\gamma)\in(0,2]$.
In the case $\zeta=0$, $\alpha_{b_t}=o(\log^\beta t)$ as $t\to\infty$.

Here is our main result. The constant $\widetilde\chi$
appearing in \eqref{mainreseq} depends only on $A$, $\gamma$ and $\kappa$ and
will be defined in Subsection~\ref{defs}.

\begin{theorem}
\label{mainres}
Let $d=1$ and suppose that Assumption~(H) and Assumption~(G) hold.
Define $t\mapsto b_t$ as in \eqref{b_t}, and let $\widetilde\chi$
be the constant in Theorem~1.5 of\/\/ \cite{BK00}. Then
\begin{equation}
\label{mainreseq}
\lim_{t\to\infty}\frac{\alpha_{b_t}^2}t\log u(t,0)=-\widetilde\chi,
\qquad \prob\text{\rm-almost surely.}
\end{equation}
\end{theorem}
\vspace{2mm}

Interestingly, if $\ell\mapsto G(\ell)$ has a power-law decay as
$\ell\to\infty$, the lower-tail dependence of the rate can
explicitly be computed. This allows for an easy comparison with
the assertion in Theorem~1.5 of \cite{BK00}. Let $t\mapsto b_t^*$
be the scale function introduced in \cite{BK00}:
\begin{equation}
\label{b_t*}
\frac{b_t^*}{\alpha_{b_t^*}^2}=\log t.
\end{equation}
Recall $\zeta_*=\sup\{\delta\ge0\colon\langle[\log(-\xi(0)\vee1)]
^\delta\rangle<\infty\}$. For $G$ decaying with a power law,
necessarily, $G(\ell)=\ell^{-\zeta_*+o(1)}$. The following is an
immediate consequence of Theorem~\ref{mainres} and the regularity
of $t\mapsto\alpha_t$:

\begin{cor}\label{special}
Let $d=1$, suppose $\prob(\xi(0)=-\infty)=0$ and suppose that
Assumption~(H) holds. Assume that either $\zeta_*=1$ or
$\zeta_*\in(0,1)$ and $G(\ell)=\ell^{-\zeta_*+o(1)}$ as
$\ell\to\infty$. Then
\begin{equation}
\label{correseq}
\lim_{t\to\infty}\frac{\alpha_{b_t^*}^2}t\log u(t,0)=
-\zeta_*^{-\beta}\widetilde\chi, \qquad \prob
\text{\rm-almost surely.}
\end{equation}
where $\beta=2(1-\gamma)/(1-3\gamma)$.
\end{cor}

\begin{remark}
\label{rem2}
By comparison of Corollary~\ref{special} and Theorem~1.5 of
\cite{BK00}, $\zeta_*=1$ is necessary  and sufficient for the
assertion of the latter to hold, at least in the class of
distribution with $G$ decaying as a positive power. In particular,
the condition that $\langle\log(-\xi(0)\vee 1)\rangle<\infty$ in
\cite{BK00} is only marginally non-optimal because Theorem~1.5
also literally holds if we just assume that $[\log(-\xi(0)\vee
1)]^\delta$ be integrable for any $\delta<1$. This answers the
first of the questions above.
\end{remark}
\begin{remark}
\label{rem2'}
The cases with $\zeta_*>0$ have a different absolute size of the
rate while the time dependence remains as for $\zeta_*=1$.
However, when $\zeta_*=0$, also the time dependence changes. For
instance, in the aforementioned example $\prob(\log(-\xi(0)\vee
1)\in dx)\sim C/[x\log^{1+\theta}(x)]\,dx$ as $x\to\infty$ (see
Remark~\ref{rem1'}), $\alpha_{b_t}^2=[\log\log t]^{\beta+o(1)}$,
which grows much slower than in the case $\zeta_*>0$. For yet
thicker lower tails, even slower growths are possible. We conclude
that the result of Theorem~1.5 of \cite{BK00} qualitatively
changes only when $-\xi(0)$ lacks all positive logarithmic
moments.
\end{remark}

The remainder of this paper is organized as follows: In the next
section we define some important objects and use them to give a
heuristic outline of the proof. The actual proof comes in
Section~\ref{proof}. Since many steps can almost literally be
taken over from \cite{BK00}, we stay as terse as possible. The
essentially novel part are Lemmas~\ref{trivbound}, \ref{LLN},
and~\ref{last}.

\section{Definitions and heuristics}

\subsection{Auxiliary objects}
\label{defs}

\noindent
For the sake of both completeness and later reference, we will now
introduce the objects needed to define the quantity
$\widetilde\chi$ in Theorem~\ref{mainres}. Then we proceed by
recalling the Feynman-Kac representation and some formulas for
Dirichlet eigenvalues.

\subsubsection{Definition of $\widetilde\chi$}
Let $\F_R$ be the set of continuous functions
$f\colon\R\to[0,\infty)$ satisfying $\supp f\subset [-R,R]$ and
having total integral equal to one. Let $C^+(R)$ (resp., $C^-(R)$)
be the set of continuous functions $[-R,R]\to [0,\infty)$ (resp.\
$[-R,R]\to (-\infty,0]$). For $H$ in the $\gamma$-class, let
$\skrih_R\colon C^+(R)\to(-\infty,0]$ be the functional defined by
\begin{equation}
\skrih_R(f)=-A\int_{[-R,R]} f^\gamma\, \1_{\{f>0\}}\, dx,
\end{equation}
where $A$ is as in \eqref{mainass}.

Let $\L_R\colon C^-(R)\to[0,\infty]$ be the Legendre transform of
$\skrih_R$:
\begin{equation}\label{calLdef}
\L_R(\psi)=\sup\bigl\{(f,\psi)-{\mathcal
H}_R(f)\colon f\in C^+(R),\,\supp f\subset\supp\psi\bigr\},
\end{equation}
where $(f,\psi)=\int f(x)\psi(x)dx$. Conventionally,
$\L_R(0)=\infty$. If $H$ is in the $\gamma$-class with a
$\gamma\in[0,1)$, $\L_R(\psi)$ can explicitly be computed: for any
$\psi\in C^-(R)$, $\psi\not\equiv0$,
\begin{equation}
\label{L_R}
\L_R(\psi)=\begin{cases}(1-\gamma^{-1})(A\gamma)^{\frac{1}{1-\gamma}}
\int |\psi(x)|^{-\frac\gamma{1-\gamma}}\,dx, \quad&\text{if }
\gamma\in(0,1),\\-A|\supp \psi|,&\text{if }\gamma=0,
\end{cases}
\end{equation}
where $|\supp \psi|$ is the Lebesgue measure of $\supp \psi$.
(Here $\L_R(\psi)=\infty$ whenever $\gamma\in(0,1)$ and the
integral diverges.)

The last object we need is the principal eigenvalue of the
operator $\kappa\Delta+\psi$ on $L^2([-R,R])$ with Dirichlet
boundary conditions:
\begin{equation}
\label{conteigen}
\lambda_R(\psi)=\sup\bigl\{(\psi, g^2)-\kappa\|\nabla
g\|_2^2\colon g\in C_{\rm c}^\infty(\supp\psi,\R),\|g\|_2
=1\bigr\},
\end{equation}
with the interpretation $\lambda_R(0)=-\infty$. Then
\begin{equation}
\label{chitildedef}
\widetilde\chi=-\sup_{R>0}\,\sup\left\{\lambda_R(\psi)\colon\psi\in
C^-(R),\, {\mathcal L}_R(\psi)\leq 1\right\}.
\end{equation}
As was proved in \cite{BK00}, $\widetilde\chi\in(0,\infty)$.

\begin{remark}
In $d=1$, the minimizer of an associated variational problem
(namely, that for the annealed or moment asymptotics) can
explicitly be computed, see \cite{BK98}. Proposition~1.4 of
\cite{BK00} then allows $\widetilde \chi$ to be evaluated in a
closed form. Except for $\gamma=0$, no such expression is known in
higher dimensions.
\end{remark}

\subsubsection{Feynman-Kac formula, Dirichlet eigenvalues}
Let $(X(s))_{s\in[0,\infty)}$ be the continuous-time simple random
walk on $\Z$ with generator $\kappa\Delta^{\rm d}$. We use $\E_x$
to denote the expectation with respect to the walk starting at
$x$. The Feynman-Kac representation for
$u(t,\cdot)$ then reads
\begin{equation}
\label{FKF}
u(t,x)=\E_x\Bigl[\exp\Bigl\{\int_0^t\xi\bigl(X(s)\bigr)ds\Bigr\}\Bigr].
\end{equation}
Given $R>0$, let $Q_R=[-R,R]\cap\Z$ and let $u_R(t,x)$ be the
solution to the system \eqref{Anderson} in $Q_R$ and Dirichlet
boundary condition $u_R(\cdot,x)=0$ for $x\not\in Q_R$. Let
$\tau_R$ be the first exit time from $Q_R$, i.e.,
$\tau_R=\inf\{s>0\colon X(s)\not\in Q_R\}$. Then
\begin{equation} \label{FKbox}
u_R(t,x)=\E_x\Bigl[\exp\Bigl\{\int_0^t\xi\bigl(X(s)\bigr)ds\Bigr\}
\1\{\tau_R>t\}\Bigr]
\end{equation}
Note that $R\mapsto u_R(t,x)$ is increasing.

In the forthcoming developments we will also need the principal
Dirichlet eigenvalue of the operator $\kappa\DeltaD+\xi$ in the
box $z+Q_R$ centered at $z$:
\begin{equation}
\label{eigenval}
\lambdaD_{z;R}(\xi)=\sup\biggl\{\,\sum_{x\in
Q_R}\xi(x+z)g(x)^2+\kappa\sum_{x\in Q_R}
g(x)[\DeltaD g](x)\colon g\in\ell^2(Q_R),\,\Vert
g\Vert_2=1\biggr\}.
\end{equation}
Note that, by the standard eigenvalue expansion (see
\cite{BK00}),
\begin{equation}
\label{ulambda}
{\rm e}_R(z)^2 e^{t\lambdaD_{z;R}(\xi)}\le u_R(t,z)\le
\#Q_R \,e^{t\lambdaD_{z;R}(\xi)},
\end{equation}
where ${\rm e}_R(\cdot)$ is the $\ell^2$-normalized principal
eigenvector in $Q_R$. In particular, the logarithmic asymptotics
of $u_R(t,z)$ and the asymptotics of $t\lambdaD_{z;R}(\xi)$
coincide provided $R=R(t)$ does not grow too fast with $t$ (which
ensures that $t\mapsto{\rm e}_{R(t)}(z)^2$ does not decay too
fast).

\subsection{Heuristic explanation}
\label{heuristics}

\noindent
As alluded to in the introduction, \eqref{mainreseq} results from
the competition of two mechanisms: (1)~searching for optimal
shapes of the potential by the walk in \eqref{FKF} and
(2)~screening off far away sites by regions of strongly negative
potential. Let us describe this interplay in detail. To avoid
cluttering of indices we often use $\alpha(b_t)$ in the place of
$\alpha_{b_t}$.

Consider a ``macrobox'' $Q_{r(t)}=[-r(t),r(t)]\cap\Z$ with
$r(t)\approx\exp[b_t\alpha(b_t)^{-2}]$, where we think of $b_t$ as
of a yet undetermined scale function. Fix $R>0$ and a shape
function $\psi\in C^-(R)$ satisfying ${\mathcal L}_R(\psi)< 1$. A
Borel-Cantelli argument shows that there exists a randomly located
microbox in $Q_{r(t)}$, with diameter $2R\alpha(b_t)$, where $\xi$
is shaped like $\psi_t(\cdot)\approx\psi(\cdot/\alpha(b_t))/
\alpha(b_t)^2$. Let us assume that $R$ and $\psi$ approximately
maximize \eqref{chitildedef}, i.e., $\lambda_R(\psi)\approx
-\widetilde \chi$. Then the dominating strategy for the walk is to
move in a short time to that favorable microbox and spend the rest
of the time until $t$ in it. The contribution coming from the long
stay in the microbox is roughly
$\exp[t\lambda_{R\alpha(b_t)}(\psi_t)]$, which can be approximated
by $\exp[t\alpha(b_t)^{-2}\lambda_R(\psi)]\approx
\exp[-t\alpha(b_t)^{-2}\widetilde\chi]$, using the scaling
properties of the Laplace operator.

The size of the macrobox is determined by the amount of mass the
walk loses on the way from the origin to the favorable microbox,
while traveling through long stretches of large negative
potential. A calculation shows that the penalty it pays is roughly
of order $\exp[-\sum_{x=1}^{r(t)} \log(-\xi(x)\vee1)]$. (An
optimal strategy is not to spend more than $(-\xi(x)\vee1)^{-1}$
time units at each site $x$ on the way.) Under our assumptions on
the lower tails of $\xi(0)$, a Borel-Cantelli argument shows that
this penalty is roughly $\exp[-G^{-1}(1/r(t))]$ where $G^{-1}$
denotes the inverse function of $G$.

As it turns out, the two mechanisms run at optimal ``speed'' when
the two exponents are roughly of the same order, i.e.,
$G^{-1}(1/r(t))\approx t\alpha(b_t)^{-2}\approx t$, because
$\alpha_{b_t}\ll t$. Recalling that
$r(t)\approx\exp[b_t\alpha(b_t)^{-2}]$, this reasoning leads to
\eqref{b_t}. A fine tuning of $r(t)$ makes the contribution from
the travel to the microbox negligible compared to the contribution
from the stay in it, i.e., we shall in fact have
$G^{-1}(1/r(t))=o(t\alpha(b_t)^{-2})$. Hence, we obtain
\eqref{mainreseq} with $\widetilde\chi$ as in \eqref{chitildedef}.

\section{Proof of Theorem~\ref{mainres}}
\label{proof}

\noindent
As in \cite{BK00}, the main result will be proved by
separately proving upper and lower bounds in \eqref{mainreseq}.
The proof of Corollary~\ref{special} comes at the very end of this section.

\subsection{The upper bound}
\label{upperbd}

\noindent
Recall  the notation of Subsection~\ref{defs}, in
particular that
$Q_R=[-R,R]\cap\Z$.
Let
\begin{equation}
\label{r(t)}
r(t)=-\frac3{G(t)}\log G(t).
\end{equation}
Note that $r(t)= t^{\zeta+o(1)}$ as $t\to\infty$
if $\widetilde G(\ell)=\ell^{-\zeta+o(1)}$ as $\ell\to\infty$.
Abbreviate $B_R(t)=Q_{r(t)+2\lfloor R\rfloor}$.

The crux of the proof of the upper bound in Theorem~\ref{mainres} is the
following
generalization of Proposition~4.4 of \cite{BK00} adapted to the
new definition of $r(t)$.

\vbox{
\begin{proposition}
\label{propcompact}
There exists a constant
$C=C(\kappa)>0$ and a random variable $C_\xi\in(0,\infty)$ such that,
$\prob$-almost
surely,
for all $R,t>C$,
\begin{equation}
\label{compact}
u(t,0)\leq C_\xi e^{-t}+e^{Ct/R^2}3r(t)
\exp\Bigl\{t\max_{z\in B_R(t)}\lambdaD_{z;2R}(\xi)\Bigr\}.
\end{equation}
\end{proposition}
}

\begin{proofsect}{Proof of Theorem~\ref{mainres}, upper bound}
With Proposition~\ref{propcompact} in the hand, the proof goes
along very much the same lines as in \cite{BK00}. Indeed, let
$r(t)$ be as in \eqref{r(t)} and set $R$ in \eqref{compact} to be
$R\alpha(Kb_t)$, where $K>0$ will be chosen later and $R$ will
tend to $\infty$. Let $H$ be in the $\gamma$-class and recall that
$\alpha_t=t^{\nu+o(1)}$ with $\nu=(1-\gamma)/(3-\gamma)$.

Abbreviate $B(t)=B_{R\alpha(Kb_t)}(t)$ and
$\lambda(z)=\lambdaD_{z;{2R\alpha(Kb_t)}}(\xi)$, and note that
$r(t)\le e^{o(t\alpha(b_t)^{-2})}$. Then, using also that
$\lim_{t\to\infty}\alpha(Kb_t)/\alpha(b_t)=K^\nu$, we have from
\eqref{compact} that
\begin{equation}
\label{upperas}
\limsup_{t\to\infty}\frac {\alpha_{b_t}^2}t\log u(t,0)\leq \frac
C{K^{2\nu} R^2}+\limsup_{t\to\infty}\Bigl[\alpha_{b_t}^2
\max_{z\in B(t)}\lambda(z)\Bigr],
\end{equation}
$\prob$-almost surely. Abbreviating $M(t)=\max_{z\in
B(t)}\lambda(z)$,
we have to prove that, for any $\eps>0$,
\begin{equation}\label{aim1}
\limsup_{t\to\infty}\alpha_{b_t}^2 M(t)\leq -\widetilde \chi+\eps,\qquad
\prob\text{-almost surely},
\end{equation}
for some appropriate $K\in(0,\infty)$ and sufficiently large $R$.

Note that the eigenvalues $\lambda(z)$ have identical distribution.
Furthermore,
their exponential moments can be estimated by
\begin{equation}\label{expmom}
\limsup_{R\to\infty}\,\limsup_{t\to\infty}\frac{\alpha_{b_t}^2}{b_t}
\log\bigl\langle e^{Kb_t\lambda(z)}\bigr\rangle\le -K^{1-2\nu}\chi,
\end{equation}
where $\chi\in(0,\infty)$ is a constant related to
$\widetilde\chi$, see \cite{BK00}. Since $t\mapsto M(t)$ is
increasing and $t\mapsto\alpha_{b_t}$ slowly varying, it suffices
to prove \eqref{aim1} for $t$ taking only a discrete set of values; the
main difference compared to \cite{BK00} is that now we take
\begin{equation}
\frac1{G(t)}\in\{e^n\colon n\in\N\}.
\end{equation}
Let $G(t)=e^{-n}$ and note that \eqref{b_t} implies that
$b_t\alpha_{b_t}^{-2}=n$. The proof now proceeds exactly as in
\cite{BK00}: We let
$p_n(\eps)=\prob(M(t)\alpha_{b_t}^2\ge-\widetilde\chi+\eps)$ and use
the Chebyshev inequality and \eqref{expmom} to derive that $p_n(\eps)$ is
summable on $n$
for all $\eps>0$, provided $K$ is chosen appropriately and $R$ is sufficiently
large. The claim
is finished using the Borel-Cantelli lemma.
\qed
\end{proofsect}

It remains to prove Proposition~\ref{propcompact}.  In
\cite{BK00}, the choice $t\log t$ for $r(t)$ allowed us to use a
simple probability estimate for the simple random walk; in
particular, the corresponding bound \eqref{compact} held true
uniformly in all non-positive potentials. In our present cases,
$r(t)$ is typically much smaller than $t\log t$ and the potential
has to cooperate to get the bound \eqref{compact}. Unlike in
\cite{BK00}, the role of the potential is actually dominant in the
cases of our present interest.

\begin{lemma}
\label{trivbound}
For any $b\in(2\kappa,\infty)$ there is a random variable
$C(\xi)\in(0,\infty)$ such that, $\prob$-almost surely,
\begin{equation}
\label{aprioribd}
u(t,0)-u_R(t,0)\le
C(\xi)\biggl(\,\prod_{x=0}^R\frac{b}{-\xi(x)\vee b}
+\prod_{x=-R}^{0}\frac{b}{-\xi(x)\vee b}\biggr),\qquad R\in\N, t\geq 0.
\end{equation}
\end{lemma}

\begin{proofsect}{Proof}
Let $(X_k)_{k\in\N_0}$ be the embedded discrete-time simple random
walk on $\Z$ and let $\ell_n(x)$ be its local times defined by
$\ell_n(x)=\sum_{k=1}^n \1\{X_k=x\}$. Let $\ED_y$ denote the
expectation with respect to the discrete-time  walk, starting at
$y\in\Z$. Abbreviate $\xi_k=\xi(X_k)$ and $\widehat
u_R(t,0)=u(t,0)-u_R(t,0)$. Then, by \eqref{FKF} and \eqref{FKbox},
\begin{equation}
\label{discreterep}
\widehat u_R(t,0)=e^{-2\kappa t}\sum_{n\ge R} (2\kappa)^{n}
\,\ED_0\biggl[\int_{\vartriangle_n\!(t)} dt_1\dots dt_{n}\,
\exp\Bigl\{\sum_{k=0}^n\xi_kt_k\Bigr\}
\1\{\supp\ell_n\not\subset Q_R\}\biggr],
\end{equation}
where
$\vartriangle_n\!\!(t)=\{(t_1,\dots,t_{n})\in(0,\infty)^n\colon
t_1+\dots+t_{n}\le t\}$, and $t_0$ is a shorthand for
$t-(t_1+\dots+t_{n})$.

Fix $b>2\kappa$ and define
\begin{equation}
\mathcal A_n=\bigl\{x\in\supp\ell_n\colon \xi(x)\le -b\bigr\}.
\end{equation}
Let
\begin{equation}
{\mathcal I}_n=\bigl\{k\in\{1,\dots, n\}\colon X_k\notin\mathcal A_n\bigr\}
\end{equation}
be the set of all the times at which the walk visits a point $x$ with $\xi(x)>
-b$.

By relaxing the constraint $t_1+\dots+t_n\le t$ in
$\vartriangle_n\!\!(t)$ to $t_k\le t$ for every $k\in {\mathcal
I}_n$, neglecting the terms with $k\in{\mathcal I}_n\cup\{0\}$ in
the exponential, and integrating out $t_1,\dots,t_{n}$, we get
\begin{equation}
\label{bd}
\widehat u_R(t,0)\le e^{-2\kappa t}\sum_{n\ge R}\sum_{m=0}^{n}
\frac{(2\kappa t)^m}{m!}\ED_0\biggl[ \1\bigl\{\# {\mathcal
I}_n=m\bigr\}\, \1\{\supp\ell_n\not\subset Q_R\}\prod_{0<k\leq
n\colon k\notin{\mathcal I}_n} \frac {2\kappa}{-\xi_k}\biggr].
\end{equation}
Neglecting the first indicator and the restriction to $m\leq n$,
we can carry out  the sum over $m$ in \eqref{bd} and find that
\begin{equation}
\widehat u_R(t,0)\leq \sum_{n\ge
R}\ED_0\biggl[\1\{\supp\ell_n\not\subset Q_R\}\prod_{x\in{\mathcal
A}_n}\Bigl(\frac{2\kappa}{-\xi(x)}\Bigr)^{\ell_n(x)}\biggr].
\end{equation}
On $\{\supp \ell_n\not\subset Q_R\}$, the walk visits either all
sites in $\{0,\dots,R\}$ or all sites  in $\{-R,\dots,0\}$. Hence,
we can estimate
\begin{equation}
\1\{\supp\ell_n\not\subset Q_R\}\prod_{x\in{\mathcal A}_n}\frac
{2\kappa}{-\xi(x)}\leq \prod_{x=1}^R\frac{b}{-\xi(x)\vee b}
+\prod_{x=-R}^{1}\frac{b}{-\xi(x)\vee b}.
\end{equation}
The claim \eqref{aprioribd}
then follows from the assertion
\begin{equation}
\label{sum}
\sum_{n=1}^\infty\,\ED_0\biggl[\,\prod_{x\in{\mathcal
A}_n}\Bigl(\frac{2\kappa}{b}\Bigr)^{\ell_n(x)-1}\biggr]<\infty
\qquad\prob\text{-almost surely,}
\end{equation}
where we used that $\xi(x)\le-b$ whenever $x\in{\mathcal A}_n$.
(The term with $x=0$ in \eqref{aprioribd} can be added or removed
at the cost of changing $C(\xi)$ by a finite amount.)

Let us prove that \eqref{sum} holds. First we note that $\mathcal
A_n$ contains in every sufficiently large interval in $\Z$ a
positive fraction of sites. Indeed, put
$p=\prob(\xi(0)>-b)\in(0,1]$ and note that by Cram\'er's theorem
we have $\prob(\#(\mathcal A_n\cap I)\le \frac {p}2\#I)\le e^{-c\#
I}$ for every bounded interval $I\subset \Z$ and some $c>0$
independent of $I$. A routine application of the Borel-Cantelli
lemma implies that
\begin{equation}\label{Abig}
\forall\,\text{interval }  I\subset[-n,n]\cap\Z\colon\qquad \#
I\geq n^{1/4}\,\Longrightarrow\, \# (\mathcal A_n\cap I)> \frac
{p}2\#I,
\end{equation}
for $n$ large enough, $\prob$-almost surely.

Now we prove that with high probability, there are sufficiently
large intervals which are traversed from one end to the other at least
twice by the random walk $(X_k)_{k=0,\dots,n}$. Fix $K_n=\lfloor
3\log n\rfloor$ and abbreviate $k_n=\lfloor n/K_n\rfloor$. We
divide the walk into $K_n$ pieces $(X^{(i)}_k)_{k=0,\dots,k_n}$
(neglecting a small overshoot) with
$X^{(i)}_k=X_{(i-1)k_n+k}-X_{(i-1)k_n}$ for $i=1,\dots,K_n$. Note
that these $K_n$ walks are independent copies of each other. Let
us introduce the events
\begin{equation}
B_n=\bigcap_{i=1}^{K_n-1}\bigl\{\sgn X_{k_n}^{(i)}=\sgn
X_{k_n}^{(i+1)}\bigr\} \qquad\mbox{and}\qquad
C_n=\bigcup_{i=1}^{K_n}\bigl\{\max_{1\le k\le k_n}
\bigl|X_k^{(i)}\bigr|\le L_n\bigr\},
\end{equation}
where $L_n=\sqrt{k_n/(\eta\log n)}$. It is elementary that
$\P_0^{\rm d}(B_n)\leq 2^{-K_n+1}\leq n^{-2+o(1)}$ as
$n\to\infty$. Furthermore, with the help of a concatenation
argument and convergence of simple random walk to Brownian motion
we derive that $\P_0^{\rm d}(C_n)\leq n^{-2+o(1)}$, whenever
$\eta>0$ is large enough. Now we estimate
\begin{equation}
\label{3.17}
\ED_0\biggl[\,\prod_{x\in{\mathcal
A}_n}\Bigl(\frac{2\kappa}{b}\Bigr)^{\ell_n(x)-1}\biggr]\leq
\P_0^{\rm d}(B_n)+ \P_0^{\rm d}(C_n)+\ED_0\biggl[\,\1_{B_n^{\rm
c}\cap C_n^{\rm c}}\prod_{x\in{\mathcal
A}_n}\Bigl(\frac{2\kappa}{b}\Bigr)^{\ell_n(x)-1}\biggr].
\end{equation}
Note that, on $B_n^{\rm c}\cap C_n^{\rm c}$, there is an interval
$I\subset[-n,n]\cap\Z$ with $\# I\geq L_n$ such that every point
of $I$ is visited by at least two of the subwalks, i.e., we have
$\ell_n(x)\geq 2$ for any $x\in I$. If $n$ is sufficiently large,
we deduce from \eqref{Abig} that there are at least $pL_n/2$
points $x$ with $\ell_n(x)\geq2$. By using this in \eqref{3.17},
we have
\begin{equation}
\ED_0\biggl[\,\prod_{x\in{\mathcal
A}_n}\Bigl(\frac{2\kappa}{b}\Bigr)^{\ell_n(x)-1}\biggr]\le
n^{-2+o(1)}+\Bigl(\frac{2\kappa}{b}\Bigr)^{L_n p/2},\qquad
n\to\infty.
\end{equation}
The right hand side is clearly summable on $n\in\N$ since $2\kappa/b<1$.
This finishes the
proof.
\qed
\end{proofsect}

Our next task is to get a good estimate on the size of the
products in \eqref{aprioribd}.

\begin{lemma}
\label{LLN}
Suppose that $\langle \log(-\xi(0)\vee1)\rangle=\infty$. Then, for
all $b\ge1$,
\begin{equation}
\label{LLNbd}
\lim_{n\to\infty}
\frac 1{G^{-1}(1/n)}\sum_{x=1}^{\lfloor 2n\log n\rfloor}
\log\Bigl(\frac{-\xi(x)\vee b}{b}\Bigr)=\infty\qquad
\prob\text{\rm -almost surely.}
\end{equation}
\end{lemma}

\begin{proofsect}{Proof}
Abbreviate $N_n=\lfloor 2n\log n\rfloor$ and let $b\ge1$. Then
\begin{equation}
\log\Bigl(\frac{-\xi(x)\vee b}{b}\Bigr)\ge\log(-\xi(x)\vee1)
-\log b.
\end{equation}
Using this estimate and the Chebyshev inequality, we have
for any $\theta>0$ that
\begin{equation}\label{esti}
\!\!\!\prob\biggl(\sum_{x=1}^{N_n} \textstyle{\log\bigl({\frac{-\xi(x)\vee
b}b}\bigr)\le \theta G^{-1}(1/n)\biggr) \le \exp\bigl\{-N_n
G(1/\lambda)+N_n\lambda\log b
+\lambda\theta G^{-1}(1/n)\bigr\}},
\end{equation}
for any $\lambda>0$. Set $\lambda=1/G^{-1}(1/n)$ and note that we
have $G(1/\lambda)/\lambda\to\infty$ as $\lambda\downarrow 0$, due
to $\langle\log(-\xi(0)\vee1)\rangle=\infty$. Consequently, the
term with $\log b$ is negligible and the right-hand side of
\eqref{esti} is bounded by $n^{-2+o(1)}$. The claim is
finished by the Borel-Cantelli lemma.
\qed
\end{proofsect}

\begin{proofsect}{Proof of Proposition~\ref{propcompact}}
Pick any $b\in(2\kappa,\infty)$. Let $t_0$ be so large such
that the sum in \eqref{LLNbd} with $n=\lceil 1/G(t)\rceil$ for all $t\geq t_0$
exceeds $G^{-1}(1/n)$. Note that $r(t)\ge\lfloor 2n\log
n\rfloor$. Combining the results of Lemma~\ref{trivbound} for
$R=r(t)$ and Lemma~\ref{LLN}, we derive, for sufficiently large $n$ resp.\
$t$, the bound
\begin{equation}
u(t,0)-u_{r(t)}(t,0)\le 2C(\xi) \exp\bigl(-G^{-1}(1/n)\bigr),
\end{equation}
where $C(\xi)$ is the constant from \eqref{aprioribd}. But
$G^{-1}(1/n)\ge t$ by our choice of $n$, which means that
$u(t,0)-u_{r(t)}(t,0)\le C_\xi e^{-t}$, where $C_\xi=2C(\xi)\vee
e^{t_0}$. The rest of the argument does not involve the particular
form of $r(t)$ and can directly be taken over from \cite{BK00}.
\qed
\end{proofsect}

\subsection{The lower bound}
\label{lowerbd}

\noindent
Unlike the upper bound, the lower bound was basically proved
already in \cite{BK00}, up to a change of the spatial scale and
Lemma~\ref{last} below. For this reason, we shall only indicate
the necessary changes.

First we prove the following converse of Lemma~\ref{LLN}:

\begin{lemma}
\label{last}
Fix $\eta\in(0,1)$ and let $\widetilde G_\eta$ satisfy
(ii) and (iii) in Assumption~(G).
Then there exists a $\varrho\in(0,\infty)$ such
that
\begin{equation}
\limsup_{n\to\infty}\frac 1{\widetilde G_\eta^{-1}(\varrho/n)}
\sum_{x=1}^n\log\bigl(-\xi(x)\vee1\bigr)\leq 1
\qquad\prob\text{\rm-almost surely.}
\end{equation}
\end{lemma}

\begin{proofsect}{Proof}
The argument is based on the asymptotic
sublinearity of $1/\widetilde G_\eta$ at infinity. However, in order to
have sublinearity on the whole interval $(0,\infty)$, we first
construct an auxiliary modification of $\widetilde G_\eta$.

Let $x_0>0$ be such that $1/\widetilde G_\eta$ is positive, increasing,
and concave on $[x_0,\infty)$. Let $D_0$ to be the right
derivative of $1/\widetilde G_\eta$ at $x_0$. Define $\widehat
G_\eta\colon(0,\infty)\to(0,\infty)$ by the formula
\begin{equation}
1/\widehat G_\eta(x)=\begin{cases} D_0 x\quad&\text{if }x\le x_0,\\
1/\widetilde G_\eta(x)+D_0 x_0-1/\widetilde G_\eta(x_0)\quad&\text{if }x>x_0.
\end{cases}
\end{equation}
Note that $1/\widehat G_\eta$ is positive, increasing, concave and
hence sublinear on $(0,\infty)$. Moreover, Assumption~(G)(iii)
holds true for $\widetilde G_\eta$ replaced by $\widehat G_\eta$.

For $a\ge 1$, abbreviate $Y_a(x)=\log(-\xi(x)\vee a)$. Choose
$a=e^{x_0}$ and estimate, for $n\to\infty$,
\begin{equation}
\label{esticonv}
\frac 1{\widetilde G_\eta\Bigl(\sum_{x=1}^n Y_a(x)\Bigr)}\leq\frac{1+o(1)}{
\widehat G_\eta\Bigl(\sum_{x=1}^n Y_a(x)\Bigr)}\le (1+o(1))
\sum_{x=1}^n \frac 1{\widehat G_\eta(Y_a(x))},
\end{equation}
where we used the fact that
$\sum_{x=1}^n Y_a(x)\to\infty$ almost surely,
and sublinearity of $1/\widehat G_\eta$. Since we have that
$\langle 1/\widehat G_\eta(Y_a(x))\rangle<\infty$, the Strong Law of
Large Numbers tells us that the right-hand side of
\eqref{esticonv} is almost surely no more than $\varrho n$, where
for $\varrho$ we can take, for instance,
\begin{equation}
\varrho = 2\bigl\langle 1/\widehat G_\eta(Y_a(0))\bigr\rangle.
\end{equation}
Hence,
we derive
\begin{equation}
\sum_{x=1}^n Y_1(x)\le \sum_{x=1}^n Y_a(x)
\le \widetilde G_\eta^{-1}(\varrho/n),
\end{equation}
which directly yields the desired claim.
\qed
\end{proofsect}

Another important ingredient is the following adaptation of the
crucial Proposition~5.1 of \cite{BK00} to the present situation.
For $\eta\in(0,1)$, choose $\varrho$ as in Lemma~\ref{last} and let
this time
\begin{equation}\label{gammatlower}
\gamma_t=\frac{\varrho}{\widetilde G_\eta(t\alpha_{b_t}^{-3})}
\end{equation}
be
the size of the macrobox $Q_{\gamma_t}$ (see
Subsection~\ref{heuristics}). Note that
$t^{\eta\zeta+o(1)}\leq\gamma_t\leq t^{\zeta+o(1)}$ as $t\to\infty$ if
$G(\ell)=\ell^{-\zeta+o(1)}$ as $\ell\to\infty$. Suppose without
loss of generality that $t\mapsto\gamma_t$ is increasing.

Define for each $\psi\in C^-(R)$ a ``microbox''
\begin{equation}
\label{Qdef}
Q^{(t)}=\begin{cases}Q_{R\alpha(b_t)}\quad&\text{if
}\gamma\not=0,\\ Q_{R\alpha(b_t)}\cap\supp\psi_t\quad&\text{if
}\gamma=0,\\
\end{cases}
\end{equation}
where $\psi_t\colon \Z\to (-\infty,0]$ is the function
$\psi_t(\cdot)=\psi(\cdot/\alpha_{b_t})/\alpha_{b_t}^2$. The
crucial input for the lower bound is the following claim, which
says that, with probability one provided $\L_R(\psi)<1$ and $t$ is
large, there is at least one microbox $Q^{(t)}$ in $Q_{\gamma_t}$,
where $\xi$ is no less than (the accordingly shifted) $\psi_t$.

\begin{proposition}
\label{ximicrobox}
Let $R>0$ and fix $\psi\in C^-(R)$ satisfying $\L_R(\psi)< 1$. Let
$\eps>0$ and suppose  Assumptions~(G) and (H) hold. Then the
following holds almost surely: For each $\eta\in(\L_R(\psi),1)$,
there is a\/ $t_0=t_0(\xi,\psi,\eps,R,\eta)<\infty$ such that for
each $t\ge t_0$, there is a\/ $y_t\in Q_{\gamma_t}$ with
\begin{equation}\label{xilowerbound}
\xi(z+y_t)\geq\psi_t(z)-\eps\alpha_{b_t}^{-2} \qquad
\forall z\in Q^{(t)}.
\end{equation}
\end{proposition}

\begin{proofsect}{Proof} 
We begin by formalizing the event in \eqref{xilowerbound}; in
order to later approximate continuous $t$ by a discrete variable,
we write $\eps/2$ instead of $\eps$:
\begin{equation}
\label{Adef}
A_y^{(t)}=\bigcap_{z\in Q^{(t)}}
\bigl\{\xi(y+z)\geq \psi_t(z)-\textstyle{\frac
{\eps}{2\alpha(b_t)^2}}\bigr\}.
\end{equation}
Note that the probability of $A_y^{(t)}$ does not depend on $y$
and note that different $A_y^{(t)}$'s are independent if the $y$'s
have distance larger than $3R\alpha(b_t)$ from each other. The
proof of Lemma~5.5 in \cite{BK00} shows that $\prob(A_0^{(t)})\geq
G(t)^{\L_R(\psi)+o(1)}$ as $t\to\infty$ (the only modification
required is to replace every occurrence of $t$ in the meaning
$\exp\{b_t\alpha(b_t)^{-2}\}$ by $1/ G(t)$).

In order to prove our claim, it is sufficient to show the
summability of
\begin{equation}
\label{ptdef}
p_t=\prob\Bigl(\bigcap_{y\in
Q_{\gamma_t}}\bigl(A_y^{(t)}\bigr)^{\rm c}\Bigr)
\end{equation}
over all $t>0$ such that $1/G(t)\in \{e^n\colon n\in\N\}$. (The
sufficiency follows from the facts that
$\alpha(b_t)/\alpha(b_{et})\to1$ as $t\to\infty$ and that
$t\mapsto b_t$ and $t\mapsto\gamma_t$ are increasing. The error
terms are absorbed into an extra $\eps/2$ in \eqref{xilowerbound}
compared to \eqref{Adef}, see \cite{BK00}.)

Using the independence of $A_y^{(t)}$ for $y\in
B(t)=Q_{\gamma_t}\cap \lfloor 3R\alpha(b_t) \rfloor\Z$ and the
bound $\prob(A_0^{(t)})\geq G(t)^{\L_R(\psi)+o(1)}$, we easily
derive
\begin{equation}
\label{ptesti}
p_t\leq
\bigl(1-G(t)^{\L_R(\psi)+o(1)}\bigr)^{\#B(t)}
\leq\exp\biggl\{-\frac{ G(t)^{\L_R(\psi)+o(1)}}{\alpha(b_t)
\widetilde G_\eta(t\alpha(b_t)^{-3})}\biggr\},
\end{equation}
where we used that $\#B(t)\ge 2\gamma_t/(3R\alpha(b_t))$ and then
applied the definition of $\gamma_t$.

Use concavity of $1/\widetilde G_\eta$ to estimate $1/\widetilde
G_\eta(t\alpha(b_t)^{-3})\geq \alpha(b_t)^{-3}/\widetilde
G_\eta(t)$ and use Assumption~(G)(i) to bound $\widetilde
G_\eta(t)$ by $G(t)^{\eta+o(1)}$. Furthermore, since $\alpha(b_t)$
is bounded from above by a positive power of
$b_t\alpha(b_t)^{-2}$, we see from \eqref{b_t} that
$\alpha(b_t)=G(t)^{o(1)}$. Applying all this reasoning  on the
right-hand side of \eqref{ptesti}, we see that
$p_t\leq\exp(-G(t)^{\L_R(\psi)-\eta+o(1)})$ as $t\to\infty$, which is
summable on the sequence of $t$ such that $1/G(t)\in\{e^n\colon
n\in \N\}$. This finishes the proof.
\qed
\end{proofsect}

Now we finish the proof of our main result.

\begin{proofsect}{Proof of Theorem~\ref{mainres}, lower bound}
Let $\eps>0$ and fix $R>0$ and $\psi\in C^-(R)$ such that
$\L_R(\psi)<1$. Let $\eta\in(\L_R(\psi),1)$, define $\gamma_t$
as in \eqref{gammatlower} and let $y_t$ be as in
Proposition~\ref{ximicrobox}; suppose $y_t\ge0$ without loss of
generality. Let $r_x=[-\xi(x)\vee1]^{-1}$. As in \cite{BK00}, the
lower bound will be obtained by restricting the walk in
\eqref{FKF} to perform the following: The walk keeps jumping
toward $y_t$, spending at most time $r_x$ at each site $x$ such
that it reaches $y_t$ before time $\gamma_t$. Then it stays at
$y_t$ until time $\gamma_t$ and then within $y_t+Q^{(t)}$ for the
remaining time $t-\gamma_t$.

Inserting this event into \eqref{FKF}
and invoking Markov property at time $\gamma_t$ we get
\begin{equation}
u(t,0)\geq \text{II}\times\text{III},
\end{equation}
where the same argument as in \cite{BK00} shows that
$\text{III}\ge e^{t\alpha(b_t)^{-2}[\lambda_R(\psi)-\eps]}$ for large $t$,
while for $\text{II}$ we have
\begin{equation}
\begin{aligned}
\text{II}=\int_{\vartriangle_{y_t}\!(\gamma_t)} &dt_0\dots
dt_{y_t-1}\, e^{-2\kappa\gamma_t}\exp\Bigl\{\sum_{k=0}^n
\xi_kt_k\Bigr\}\prod_{x=0}^{y_t-1}\1\{t_x\le r_{x-1}\} \\
&\ge e^{-2\kappa\gamma_t}\prod_{x=0}^{y_t-1}\Bigl[r_xe^{r_x\xi(x)}\Bigr]\ge
e^{-(2\kappa+1)\gamma_t}\exp\Bigl\{-\sum_{x=0}^{y_t-1}
\log\bigl(-\xi(x)\vee1\bigr)\Bigr\},
\end{aligned}
\end{equation}
where we recalled the notation of \eqref{discreterep}. Now
$y_t\le\gamma_t$, so using Lemma~\ref{last} we have that
\begin{equation}
\text{II}\ge e^{-(2\kappa+1)\gamma_t}\exp\bigl\{- \widetilde
G_\eta^{-1}(\varrho/\gamma_t)(1+o(1))\bigr\}=e^{-(2\kappa+1)\gamma_t-
t\alpha(b_t)^{-3}(1+o(1))},
\end{equation}
where we used the definition of $\gamma_t$. Since $1/\widetilde
G_\eta$ is asymptotically concave, $\gamma_t=\varrho/\widetilde
G_\eta(t\alpha_{b_t}^{-3})\le O(t\alpha_{b_t}^{-3})$ and the
exponent is $o(t\alpha_{b_t}^{-2})$. Consequently,
\begin{equation}
u(t,0)\ge
e^{t\alpha(b_t)^{-2}[\lambda_R(\psi)-\eps+o(1)]},
\end{equation}
where $o(1)$ still depends on $\eta$. The proof is finished by
letting $t\to\infty$ (which eliminates the dependence on $\eta$),
optimizing over $\psi$ and $R$ with $\L_R(\psi)<1$ and letting
$\eps\downarrow0$.
\qed
\end{proofsect}

\section*{Acknowledgments}
\nopagebreak
M.B. would like to thank Yimin Xiao and Oded Schramm
for discussions about the behavior of sums of i.i.d.\ random
variables with infinite mean.
\vspace{2mm}


\end{document}